\documentclass[11pt,twoside]{article}
\usepackage{macroaaa-eng}

\usepackage{latexsym}
\usepackage{verbatim}

\newcommand{\psim}{\lower.5ex\hbox{$\; \buildrel \propto \over\sim \;$}}
\newcommand{\lesssim}{\lower.5ex\hbox{$\; \buildrel < \over\sim \;$}}
\newcommand{\gtrsim}{\lower.5ex\hbox{$\; \buildrel > \over\sim \;$}}

\newcommand{\Sc}{{\cal S}}
\newcommand{\Ec }{{\cal E}}

\newcommand{\e}{\epsilon}
\newcommand{\g}{\gamma}

\newcommand{\gp}{\gamma^\prime}

\newcommand{\sT}{\sigma_{\rm T}}

\newcommand{\el}{\ell_{\rm S}}

\newcommand{\dD}{\delta_{\rm D}}

\newcommand{\iq}{\lower.5ex\hbox{$\; \buildrel \rightarrow \over {a\ll 1} \;$}}

\newcommand{\tp}{t^\prime}
\newcommand{\ep}{\epsilon^\prime}

\newcommand{\app}{{Astroparticle~Physics}}

\newcommand{\rmp}{{Reviews~of~Modern~Physics}}
\newcommand{\nat}{{Nature}}
\newcommand{\lsim}{\lower.5ex\hbox{$\; \buildrel < \over\sim \;$}}
\newcommand{\gsim}{\lower.5ex\hbox{$\; \buildrel > \over\sim \;$}}

\begin{document}

\vskip 1.0cm
\markboth{Dermer \& Fryer}{La Plata Lectures on GRBs and Fermi/GLAST}
\pagestyle{myheadings}

\vspace*{0.5cm}
\title{Gamma Ray Bursts and the Fermi Gamma Ray Space Telescope: 
Notes to the La Plata Lectures}

\author{Charles D.\ Dermer$^{1}$ \& Chris L.\ Fryer$^{2}$}
\affil{$^1$Space Science Division, Code 7650, U.\ S.\ Naval 
Research Laboratory, 4555 Overlook Ave.\ SW, Washington, DC, 
20375-5352 USA; charles.dermer@nrl.navy.mil}

\affil{$^2$Theoretical Astrophysics, Los Alamos National Laboratories, 
Los Alamos, NM 87545 USA; fryer@lanl.gov}

\begin{abstract} Gamma-ray bursts (GRBs) are a mixed class 
of sources consisting of, at least, the long duration and 
short-hard subclasses, the X-ray flashes, and the low-luminosity 
GRBs. In all cases, the release of enormous amounts of energy 
on a short timescale makes an energetic, relativistic or mildly 
relativistic fireball that expands until it
reaches a coasting Lorentz factor determined by the amount of baryons
mixed into the fireball. 
Radiation is produced when the blast wave interacts with the surrounding
medium at an external shock, or when shell collisions dissipate 
kinetic energy at internal shocks.
This series of notes is organized as follows: (1) The observational 
situation of GRBs is summarized; (2) Progenitor models of 
GRBs are described; (3) An overview of the
the blast-wave physics used to model
leptonic emissions is given; 
(4) GRB physics is applied to hadronic acceleration and 
ultra-high energy cosmic ray production; (5) Prospects 
for GRB physics and $\gamma$-ray astronomy
with the Fermi Gamma-ray Space Telescope (FGST, formerly GLAST), and
space-based and ground-based observatories
are considered. Also included are exercises and problems.

\end{abstract}
\vskip0.5in
 
\noindent{\bf 0. Introduction}
\vskip0.1in
GRBs are brief flashes of radiation at hard X-ray and soft 
$\gamma$-ray energies that display a wide variety of 
time histories.  GRBs were first detected with the Vela series of spacecraft 
at soft $\gamma$-ray energies with wide field-of-view
instruments used to monitor terrestrial nuclear explosions. 
The Burst and Transient Source Experiment (BATSE) on the {\it 
Compton Gamma Ray Observatory} showed that GRBs are not 
a Galactic disk population. The discovery
of X-ray afterglows with Beppo-SAX allowed for counterpart 
identification and redshift measurement, revealing the distance scale to 
long duration GRBs. HETE-II and Swift observations have
revealed the counterparts to the short-hard class of GRBs, and 
Swift observations give us a crucial look at the early afterglow 
X-ray behavior of GRBs.

This series of five lectures is organized as follows:
\begin{enumerate}
\item {\it Observations of GRBs.} Introduction to GRB observations;
\item {\it GRB Progenitor Models.} A brief review of GRB progenitor models; 
\item {\it Leptonic processes in GRBs--–prompt and 
afterglow emissions.} A description of the 
blast wave physics used to model leptonic emissions, 
including nonthermal synchrotron and 
synchrotron self-Compton radiations;
\item {\it Hadronic processes and cosmic rays from GRBs.} Application of blast wave physics to hadronic emissions and the acceleration of cosmic rays;
\item {\it GRB and $\g $-ray studies with the Fermi Gamma-ray Space Telescope, 
formerly the Gamma-ray Large Area Space Telescope, GLAST.} The future of GRB studies and prospects for $\gamma$-ray astronomy in light of Fermi, ground-based 
$\gamma$-ray telescopes, and complementary multiwavelength
and multimessenger observatories.
\end{enumerate}


The term ``gamma-ray bursts" is now understood to comprise several classes of these sources, 
including
\begin{enumerate}
\item Long-duration GRBs, the 
source class commonly meant by classical GRBs, which are associated
with high-mass stars and star-forming galaxies;
\item The short hard class of GRBs, which are widely thought to originate from the coalescence of compact objects;
\item The X-ray flashes, which are distinguished from the long-duration GRBs by having the peaks
of their energy output at X-ray rather than soft $\gamma$-ray energies; and 
\item Low luminosity GRBs, which produce unusually low energy releases compared to the long-duration GRBs.
\end{enumerate}
In addition to these subclasses of GRBs are the soft gamma repeaters (SGRs), which 
technically are not GRBs, but rather
related to phenomena on highly magnetized neutron stars.
Even without black holes they may have, like
radio-loud subclasses of black-hole sources, relativistic outflows.

The physics developed here can be applied to 
all these classes of GRBs, and other important cosmic
phenomena (e.g., blazars, microquasars), because all
involve the release of a large quantity of energy 
during a catastrophic event, thought to be driven by matter
accretion onto a black hole or, in the case of SGRs, a crustal
anomaly on a highly magnetized neutron star. 

\section{Gamma-ray Bursts: Overview of the Observations}

A GRB may flare up from any direction in space. The
classical, long-duration GRB (LGRB) releases most of
its energy  in hard X-ray and soft $\gamma$-ray (X/$\gamma$) energies during (to the observer) a fraction of 
a second to tens of seconds. 
There is no compelling evidence that LGRBs are recurrent events. 
Therefore, a wide
field-of-view instrument 
is necessary for serendipitous detection. 

\subsection{Discovery of Gamma Ray Bursts}

GRBs were discovered in data returned between 1967 
and 1973 by the Vela series of satellites used to monitor compliance with the nuclear test ban treaty. 
The Vela spacecraft carried non-imaging CsI detectors that were sensitive in the $\approx 200$ keV -- 1 MeV range (Klebesadel, Strong, \& Olson 1973). Above the large background, coincident events were identified in the light curves. Timing studies and triangulation were used to give an approximate direction to the GRBs, revealing a cosmic/non-terrestrial origin. [{\it Exercise:} Perform a simple timing analysis from synthetic 
satellite data to show how to reconstruct arrival direction
information. Give uncertainty analysis.]

The basic observational data in GRB studies are the spectral photon fluxes $\phi(\e;t)$ measured at time $t$ and at photon energy $h\nu = m_ec^2 \epsilon$. From this quantity, one can derive, after subtracting backgound flux, the $\nu F_\nu$ flux (cgs units of ergs cm$^{-2}$ s$^{-1}$) $f_\e(t) = m_e c^2 \int_{\e_1}^{\e_2} d\e \; \e\; \phi(\e;t) \approx m_ec^2 \e_d^2 \phi(\e_d;t)$, where $\e_d$ is the typical photon energy at which the detector is most sensitive. The fluence between times $t_1$ and $t_2$ and within the energy range $\e_1$ and $\e_2$ (cgs units of ergs cm$^{-2}$) is given by ${\cal F}(t_1,t_2,\e_1,\e_2) = m_e c^2 \int_{t_1}^{t_2}dt\;\int_{\e_1}^{\e_2} d\e \; \e\; \phi(\e;t)$. Flux and fluence distributions can be constructed from observations of many GRBs.

\subsection{BATSE Observations: GRBs are Cosmological}

The Burst and Transient Source Experiment BATSE
on CGRO consisted of an array of large area 
detectors (LADs) most sensitive in the 50-300 keV band, 
in addition to smaller spectroscopy detectors.
The BATSE has given the most extensive data 
base of GRB observations during the prompt phase. It searched for GRBs by
examining strings of data for $> 5.5\sigma$ enhancements above background on the 64 ms, 256 ms, and 1024 ms time scales, 
and triggers on GRBs as faint as $\approx 0.5$ ph cm$^{-2}$ s$^{-1}$,
corresponding to energy flux sensitivities $\lesssim 10^{-7}$ ergs cm$^{-2}$ s$^{-1}$.
At hard X-ray and soft $\gamma$-ray energies, 
the peak flux may reach hundreds 
of photons cm$^{-2}$ s$^{-1}$ in rare cases. Empirical morphological studies of LGRBs give various
phenomenological relations, including
 hardness-intensity correlation, generic hard-to-soft
evolution of $\e_{pk}(t)$, and variability-distance correlation.

Expressing the sensitivity of a high-energy radiation detector in terms of a
threshold energy flux $\Phi_{thr}$ (same units as $f_\e$) 
imposes the condition that $\Phi \geq
\Phi_{thr}$. For unbeamed sources with luminosity
$L_*$ and distance $d$, 
$\Phi = L_*/4\pi d^2$, and the maximum source distance for a give
source flux $\Phi$ is 
$$d(\Phi) = \sqrt{L_*\over 4\pi\Phi}\;.$$
(The luminosity distance which includes cosmological 
effects is defined by $d_L = \sqrt {L_*/4\pi \Phi}$, and $d\cong d_L$ at low redshifts $z \ll 1$. [{\it Exercise.}
Relate energy to fluence, including redshift.])
Hence the well-known $-3/2$ result
for sources uniformly distributed with density $n_0$ in Euclidean space,
namely
\begin{equation}
N(>\Phi) = N(< d) = 4\pi n_0 \int_0^{d(\Phi)}dx\;x^2  
\propto \Phi^{-3/2}\;\;.
\label{Nphi}
\end{equation}

The $\langle V/V_{max}\rangle$ statistic 
\begin{equation}
\big\langle V/V_{max}\big\rangle\;=\;{1\over N}\sum_{i=1}^N\;
\big({\Phi_i\over \Phi_{thr}}\big)^{-3/2}\;
\label{vovervmax}
\end{equation}
expresses the deviation from 0.5 expected for a uniform Euclidean
distribution of sources (Schmidt 1968). Here $V$ stands for volume, and $V_{max}$ is the 
maximum volume from which a source with flux $\Phi$ could be detected.
Values of $\langle V/V_{max}\rangle > 0.5$
represent positive evolution of sources, that is, either more sources
and/or brighter sources at large distances or earlier times.  Values of $\langle
V/V_{max}\rangle < 0.5$ represent negative source evolution, i.e.,
fewer or dimmer sources in the past. Detailed treatments of the statistical 
properties of black hole sources must consider cosmological effects and 
evolution of source properties.

The integral size distribution of BATSE GRBs in terms of peak flux $\phi_p$ is very flat below $\sim 3$ ph cm$^{-2}$ s$^{-1}$, and becomes steeper than the $-3/2$ behavior expected from a Euclidean distribution of sources at $\phi_p \gtrsim 10$ ph cm$^{-2}$ s$^{-1}$.
The directions to the BATSE GRBs are isotropically distributed in the sky and display no clustering toward the Galactic plane. When coupled with the flattening of the peak flux distribution, the implication is that we are at the center of an isotropic though bounded distribution of GRB sources. A cosmological distribution of sources is most compatible with these observations.

The duration of a GRB is defined by the time during which the middle 50\% ($t_{50}$),  90\% ($t_{90}$), or $i$\% ($\langle t_i\rangle $) of the counts 
above background are measured. 
A bimodal duration distribution is measured, irrespective of whether the 
$t_{50}$ or $t_{90}$ durations are considered
(Kouveliotou et al.\ 1993)
About two-thirds of BATSE GRBs are long-duration GRBs 
with $t_{90}\gtrsim 2$ s, with the remainder comprising the short duration GRBs. The short duration GRBs tend to have harder spectra, so that they are referred to as the short, hard class of GRBs. They are also much weaker in 
average fluence in the BATSE range. 

GRBs typically show a very hard spectrum in the hard X-ray to soft $\gamma$-ray regime, with a photon index breaking from $\approx -1$  at photon energies $E_{ph}\lesssim 50$ keV to a $-2$ to $-3$  spectrum at $E_{ph} \gtrsim$ several hundred keV. In BATSE
studies,  the
 distribution of the peak photon energies $E_{pk}$ of the time-averaged
$\nu F_\nu$ spectra of BATSE GRBs are typically found in the  100 keV - several MeV range. The time-averaged or, for very bright 
GRBs, time-sliced GRB spectrum is usually well-described
by the ``Band function" $N_B(\e)$ (Band et al.\ 1993),  
a power-law times an
exponential that smoothly connects to a steeper
power-law,
given by
$$ N_{_B}(\epsilon) = k_{_B} \; \epsilon^{\alpha} \exp[-\epsilon
(\alpha-\beta)/\epsilon_{_{\rm br}}] \; H(\epsilon; \e^{\rm B}_{\rm
min},\epsilon_{_{\rm br}})
$$
\begin{equation}
+ \;  k_{_B} \; \epsilon_{_{\rm br}}^{\alpha-\beta}
\exp(\beta-\alpha)
\; \epsilon^\beta \; H(\e;\e_{\rm br},\e^{\rm B}_{\rm max}) \; .%
\label{Nband-eq} %
\end{equation}
Here $\alpha$ and $\beta$ are the low and high energy 
photon number indices, and
$E_{\rm br} = m_ec^2\epsilon_{_{\rm br}}$ is the ``break energy.''
The Heaviside function $H(x;y,z)$ vanishes everywhere except at
$y\leq x<z$, where it equals unity. The term $k_{_B}$ is the
constant normalizing the number fluence to the $> 20$ keV BATSE
energy fluence $\Phi_{_B} (> 20 {\rm \; keV})$ of a particular
GRB [{\it Exercise:} Derive the form of the Band function
and normalizing constant, and convolve with nontrivial model
detector response.] 
Typical values of Band alphas $\alpha \approx -1$ 
and Band betas $\beta \approx -2.2$ -- $-2.5$. Deviations of these
values give valuable information about radiation processes 
and existence of separate radiative components.

\subsection{Beppo-SAX and the Afterglow Revolution}

Beppo-SAX GRB observations reveal that essentially all
long-duration GRBs have fading X-ray afterglows.
Beppo-SAX, launched April 30, 1996, carried three instruments. The Gamma Ray Burst Monitor was sensitive in the range 60 -- 600 keV to GRBs brighter than $\approx 10^{-6}$ ergs cm$^{-2}$ s$^{-1}$. The Wide Field Camera was sensitive in the range 2 -- 30 keV down to $\approx 10^{-10}$ ergs cm$^{-2}$ s$^{-1}$ and provided $\lesssim 10^\prime$ error boxes. 
The spacecraft was then 
slewed, requiring 6 -- 8 hours, but was fast enough for the Narrow Field Instruments, sensitive  to $> 0.1$ keV GRB emissions as faint as $\sim 10^{-14}$ ergs cm$^{-2}$ s$^{-1}$, to give error boxes $\lesssim 0.5^\prime$. 
The first X-ray afterglow was obtained from GRB 970228 (Costa 1997), which revealed
an X-ray source that decayed according to a power-law, $\phi_X(t) \propto t^{\chi}$, 
with $\chi \sim -1.33$. Typically, $\chi \sim -1.1$ to $-1.5$ in Beppo-SAX 
$\sim 2$ -- 10 keV X-ray studies. 

The small X-ray error boxes allowed deep optical and radio follow-up studies. 
GRB 970228 was the first GRB from which an optical counterpart was observed (van Paradijs
et al.\ 1997), and GRB 970508 was the first GRB for which a redshift was measured.
Redshifts are provided by detection of optical emission lines from the host galaxy and absorption lines in the fading optical afterglow due to the 
presence of intervening gas. 
Host galaxies of long duration GRBs are bluish star-forming galaxies, primarily consisting of dwarf irregular and spiral galaxies. No optical counterparts were detected from approximately one-half of Beppo-Sax GRBs with well-localized X-ray afterglows, and are termed ``dark" bursts. These sources may be undetected in the optical band because of 
dusty media,  
or intrinsically faint afterglows. 
These results give compelling evidence that LGRBs are associated
with star-forming galaxies 
and the deaths of massive stars,
especially given the detection of supernova emissions a few weeks
after the GRB in a few, nearby faint GRBs 
which, however, 
may not be fully representative of the LGRBs (but rather the 
low luminosity GRBs, LLGRBs. 

Apparent isotropic energy releases of LGRBs are enormous, exceeding
$10^{54}$ ergs, and the redshift distribution of Beppo-SAX, BATSE, 
HETE-II and INTEGRAL (pre-Swift) GRBs is peaked near $\langle z \rangle \approx 1$. 
Achromatic breaks in the optical curves of GRB afterglows
gives evidence for a beamed/jetted geometries, reducing the apparent 
energy release to a beaming-corrected energy release by a beaming factor 
${\cal F}_{bm}$. 
Approximately 40\% of GRBs have radio counterparts, and the transition from a 
scintillating to smooth behavior in the radio afterglow of GRB 980425 provides evidence for an expanding source.  
For BATSE/Beppo-SAX type
GRBs, the lion's share $\sim 65$\% of the energy is released in the 
form of $>25$ keV X-rays and soft $\gamma$ rays,  $\sim 7$\% is softer X-rays,
$\sim 0.1$\% in optical, during the prompt phase $t \lesssim 2\langle t_i\rangle$.

A class of X-ray rich GRBs, with durations on the order of seconds 
to minutes and X-ray fluxes in the range $10^{-8}$ -- $10^{-7}$ ergs cm$^{-2}$ s$^{-1}$ in the 2-25 keV band, was detected with many X-ray satellites, including Ariel V, HEAO-1, ROSAT, and Ginga, but conclusively established with Beppo-SAX (Heise et al.\ 2001). 
These X-ray flashes (XRFs) are ``$\gamma$-ray 
challenged," as indicated by Band-function model fits to the prompt emission spectrum.
 Several phenomenological correlations of
XRFs and LGRBs have been reported. One
goal is to establish a pseudo-redshift indicator, 
another a ``pulse" paradigm, 
and 
 correlations between the duration of quiescent and subsequent pulse periods in separated pulses. 
A quantitative relation between integrated fluence and $E_{pk}$ in well-defined pulses is reported. 

The Amati relation (Amati et al.\ 2002) correlates the $\nu F_\nu$ peak 
photon energy $E_{pk}$  with apparent
isotropic energy release ${\cal E}_{iso}= 10^{54}{\cal E}_{54}$ ergs according
to 
\begin{equation}
E_{pk} \propto {\cal E}_{iso}^{1/2}\;.
\label{Amati}
\end{equation}  
The Ghirlanda relation (Ghirlanda et al.\ 2004) correlates $E_{pk}$ with the 
collimation-corrected absolute $X/\gamma$ 
energy release ${\cal E}_{abs} = 10^{51}{\cal E}_{51}$ ergs  according to 
\begin{equation}
E_{pk} \propto {\cal E}_{abs}^{0.7}\;.
\label{Ghirlanda}
\end{equation}
The Amati and Ghirlanda relations 
are challenging to explain 
and potentially useful for 
cosmolgical studies.

\subsection{Swift and Various Classes of Bursting Sources}

The Swift Observatory is a 
NASA-ASI--supported MidEx launched 
November 20, 2004. Its main scientific payload consists of 3 instruments: the Burst Alert Telescope (BAT), triggering between 
15 and 150 keV, and a very sensitive X-ray Telescope
(XRT) operating between $\sim 0.3$ -- few keV, and 
the Ultra-violet optical telescope (UVOT). The spacecraft slews autonomously 
in 20 -- 75 s in response to triggers from the BAT.

\subsubsection{Long duration GRBs (LGRB).}

The redshift distribution of the GRBs detected with Swift,
with average Swift GRB redshift $\langle z \rangle \approx 2$,
 differing markedly from $\langle z \rangle \approx 1$ for pre-Swift GRBs. This can be explained by the different triggering energy
range of BAT vs.\ BATSE, and $E_{pk}$-flux correlations.
Knowledge of the early X-ray afterglow phase, an important 
goal realized by the Swift mission, provided surprising unpredicted behavior,
most notably rapid X-ray declines and X-ray flares to late times (Zhang et al.\ 2006;
Nousek et al.\ 2006).
The physical meaning of the rapid declines in the X-ray flux between
$\sim 100$ -- $10^3$ s is in dispute, including muti-component jet 
models, refreshed jets, and hadronic signatures.

\subsubsection{Short Hard GRBs (SGRB).}

As expected for an old population of host galaxy
progenitors, as 
would be the case for coalescing neutron stars and black holes, 
SGRBs are expected to be associated with ellipticals as well
as spirals. 
No evidence for supernova emissions
has been found in SGRB afterglows, meaning that the progenitors
are not associated with a young, massive stellar population. 
That this expectation received spectacular 
confirmation with Swift, and also HETE-II,  only opens up the surprises, including
delayed X-ray afterglow emissions, rapid X-ray declines, and 
X-ray flares.

The heterogeneous Swift/HETE sample has a redshift
distribution broadly distributed around $\langle z \rangle \approx 0.4$, and differs in important ways from the LGRBs in terms of 
host galaxies, offsets from host galaxies, 
apparent energy releases, and 
lag-luminosity relation. Absolute energy determination for SGRBs is
compromised by the difficulty of finding beaming breaks in  
light curves of SGRBs. ({Exercise: Beppo-SAX was not sensitive
to SGRBs. Why?)

\subsubsection{Low Luminosity GRBs (LLGRBs).}

The LLGRB class is the long tail to the LGRBs
to low apparent energy release (reviewed by Zhang 2007). 
LLGRBs could very well be 
a separate and distinct population from LGRBs, 
possibly associated
with magnetar activity, as indicated by 
GRB 060218, or extended black-hole fueling activity.
A possible new class of GRBs was found in relation to 
nearby GRB 060614 with no supernova emissions, though 
it could possibly be a nearby SGRB. 
Identification of LLGRBs as a separate
class, known since GRB 980425/SN 1998bw, 
makes us rethink conclusions about the relation 
of LGRBs to supernovae (SNe), which mostly derived from nearby LLGRBs.
({\it Exercise:} Report on GRB 98025.)

\subsubsection{Soft Gamma Repeaters (SGRs).}

The giant flare of December 27, 2004 
was the most intense cosmic transient 
observed historically. It was 
detected by over 20 spacecraft 
from the Earth to Saturn, and apparently released
 $\approx 1000$ times more energy than all the Milky Way's 
stars, $\gtrsim 1$ erg cm$^{-2}$ in hard X-rays and $\gamma$ rays at Earth 
(Hurley et al.\ 2005).
({\it Exercise.}
Estimate the bolometric radiation flux from the stars 
in the Milky Way. Estimate 
the total stellar energy flux of the universe.)
The event, releasing $\sim 10^{47}$ 
ergs, 
lowered the level of the Earth's ionosphere.

SGR 041227 began with an $\sim 0.2$ s long, hard spectrum spikes with $E\sim 10^{46}$ -- $10^{47}$ erg.
The spike is followed by a pulsating tail with 
$\approx 1/1000^{th}$ of the energy. Viewed from a large distance, 
only the initial spike would be visible, and 
would resemble a short GRB.
It could be detected out to 100 Mpc
GRB 050906 at $z=0.03$ could be a magnetar flare.
Giant flares like this must occur in other galaxies, and
could comprise $\sim 10$\% of the the SGRB population.

\subsection{
Summary}

\begin{enumerate}
\item Pioneering phase (1967 –- 1991): era of confusion 
\item BATSE/CGRO era (1991 --2000): GRBs are cosmological 
\item Beppo-SAX afterglow era (1996 -- 2006): distance and energy scale for classical LGRBs   (HETE-II/INTEGRAL)
\item XRFs, first seen with Ginga, clarified with Beppo-SAX
\item Low luminosity GRBs class (first example in 1998: GRB 980425/SN 1998bw)
\item Swift era (2004 -- ): early afterglows and the distance and energy scale for short-hard SGRBs (HETE-II)
\item Fermi GBM (GLAST Burst Monitor)/Fermi LAT (Large Area Telescope) era (2008 -- )
\end{enumerate}

\section{GRB Progenitor Models} 

Summarizing the observations of LGRBs, we find that
a ``typical" long-duration GRB lasts $\approx 20$ s in hard X-ray/soft $\gamma$ ray emission from keV to MeV energies. 
It takes place in star-forming (spiral or dwarf irregular) galaxies, but not in ellipticals.
It takes place in a galaxy at $\langle z \rangle \approx 1$ -- 2, and releases 
$\sim 10^{51}$  -– $10^{54}$ ergs of apparent isotropic energy in bursts of 
radiation with apparent isotropic luminosities of $\approx 10^{50}$  
-– $10^{52}$ ergs/s.
It is followed by long-lived X-ray, optical, and radio afterglow emission. 
Variability times are as short as ms 
(though more typically 1 s). How to explain this phenomenology?
The consensus of most GRB scientists is that a LGRB is 
the consequence of the collapse of a massive star 
$\approx 30 M_\odot$ to a black hole fed by an accretion torus.

In this second Lecture, we present a description of  
\begin{enumerate}
\item GRB Source Models: Core Collapse vs.\ Coalescence
\item Energy source: nuclear, gravitational, rotation
\item Jet formation: neutrinos vs.\ magnetic field energy
\item Pathways to progenitor formation of collapsars
\item Pathways to progenitor formation of merging compact objects
\end{enumerate}

\subsection{Beaming and the Fireball Model}

A strongly collimated, jetted emission helps answer the question, 
``How is it possible for a source to produce $10^{54}$ ergs of energy?" 
The answer is, you produce only a fraction of this energy, 
in beams with $\partial E/\partial \Omega \approx {\cal E}/4\pi$
in a small solid angle element $\Delta \Omega$. 
Beaming into a cone of opening half-angle 
$\theta_j \approx 5$ -- 10$^\circ$ 
decreases the gamma-ray energies by $\sim 2$  orders of 
magnitude, to $\sim 10^{51}$ erg. [{\it Exercise:} Work out the beaming
factor for a two-sided top-hat jet, and for a non-trivial jet profile.]
Optical beaming breaks are derived by equating
  the characteristic beaming angle of emission, $\approx 1/\Gamma$,
with the jet opening angle, $\theta_j$. A softening of the power
law takes place when $\theta_j \lesssim 1/\Gamma$ as deceleration of 
the blast wave reduces the $\Gamma$ factor.
Beaming also increases the total burst rate by the same factor, but this does not contradict anything we know about star formation and evolution for ${\cal F}_{bm} \sim 10^1$ -- $10^3$. 

But how to obtain large energy releases, large luminosities, and short variability time scales?, and

How does the kinetic energy of the ejecta get converted to electromagnetic radiation? 

These questions are answered by the fireball/blast wave model considered
in the next lecture. In this lecture, we instead
consider the question of the progenitor
to the $\gamma$-ray burst. 

Compared to SNe, which occur $\approx$ once per sec throughout the universe, LGRBs take place, depending on 
beaming factor, $\sim {\cal F}_{bm}$ per day $\approx 10^2 ({\cal F}_{bm}/100)$ per day throughout the universe, representing as few as 0.1\% of the SN population. 
GRB-hosted SNe, for example, SN 2002ap, tend to 
by hyper-energetic compared to ``normal" SNe. Thus Paczynski (1998) coined
the name ``hypernovae," which has 
proven to be a useful concept.

\subsection{Classes of $\gamma$ Ray Transients}

\subsubsection{Short GRBs.}

What are the mechanisms for the SGRBs? Most 
GRB scientists support a picture 
involving the merger of a compact binary system, 
such as two neutron stars, 
or a neutron star and a black hole (Eichler et al.\ 1989).
 Lack of optical and radio afterglows is explained by tenuous ISM, 
if the merger takes place outside the host galaxy. 
The class of GRBs
involving merger events, namely the SGRBs, are different in
significant ways, clearly revealing the two as having distinct
origins. But there is also another distinct class.

\subsubsection{X-Ray Flashes.}

[{\it Exercise:} Report on the properties of the XRFs.]

If XRFs are another manifestation 
of long GRBs, then are they
\begin{enumerate}
\item GRBs at high redshift?
\item GRBs observed away from the jet axis?
\item Explosions with less relativistic ejecta? 
\end{enumerate}

We have fairly complete data on one XRF (XRF020903, $z=0.251$); in this case, the answer is compatible with 3 (Soderberg et al.\ 2004). The
Amati relation gives evidence that XRFs and GRBs are 
part of the same family, consistent with a mass-loaded fireball model. 

\subsubsection{Soft Gamma Repeaters.}

What are the SGRs?  Most GRB
scientists think that the 3 or 4 SGRs in the Milky Way and the 
SGR in the LMC originate from a transient release of 
magnetic field energy in highly magnetized neutron stars 
($B > B_{cr} = m_e^2 c^3/e\hbar$). The energy 
could be released from global crustal fracture 
through B field annihilation, or when 
the magnetosphere fills with hot e$^-$e$^+$ plasma.

\subsubsection{Long duration GRBs.}

LGRB progenitors generally involve  black-hole/accretion-disk models.
To answer the energy generation and release question, 
a number of problems must be solved, including the problems of the
\begin{itemize}
\item Black Hole/Accretion disk (BHAD) + Jets;
\item Structure of a Relativistic Disk;
\item Neutrino Driven Explosions from a BHAD system; and
\item Magnetic Field Driven Explosions from a BHAD system.
\end{itemize}
To firmly resolve the astrophysical origins of GRBs, it is 
necessary to also understand the connections of the
various classes of GRBs to their stellar evolutionary past.

\subsection{Energy Sources and Energy Transport}

Three main energy sources in astrophysics are 
\begin{enumerate}
\item    Nuclear Energy
\item   Gravitational Potential Energy
\item  Black Hole Rotational Energy
\end{enumerate}
The proton-proton chain efficiency is
$0.0067$, too small to realistically power
most GRB radiations.
Gravitational energy will be released during
core collapse or coalescence event, and such 
events could also trigger a rapid discharge of rotational
energy from a rapidly spinning black hole.
Which of the other two are most important for 
GRB energy release, or are they both?

Energy in the form of photons is
transported very differently through stellar interiors
than the energy of neutrinos, 
the former dominated by absorption and reemission, and
the latter often following a rectilinear trajectory out from 
the stellar core. Simulations of core collapse give us
ways to follow different modes of energy release, also 
including gravitational radiation.

\subsection{Principal GRB Models}

The majority GRB establishment has settled on a
standard scenario for the production of 
LGRBs and SGRBs, both involving black-hole accretion disk (BHAD) 
models. The energy 
release in LGRBs is believed to result from 
the accretion of a massive torus onto a 
black hole following the collapse of the core
of a massive star to a black hole. This is 
the collapsar scenario, pioneered by Stan Woosley (1993).
Compact object mergers have a different family history,
but also end up with the release of energy by jetted
outflows from BHAD-type systems. 

A variety of scenarios for catrastophic events involving compact
stars can be imagined, for example,
\begin{enumerate}
\item Neutron Star -– Neutron star mergers, 
e.g., the Hulse-Taylor binary pulsar system
\item Black Hole –- Neutron star mergers
\item Black Hole –- White dwarf mergers
\item Collapse of the core of a rotating massive star (binary or single star)
\item Neutron star -- black hole merger with the helium core of an evolved primary, a "He-star merger"
\end{enumerate}

The collapse of the core of a rotating massive star down to a black
hole is the original scenario behind the collapsar model.  MacFadyen
\& Woosley (1999) modeled the toy case of such a collapse, placing an
artificial angular momentum profile onto a massive star whose core was
assumed to immediately collapse to a black hole.  They identified many
of the key ingredients for a succesful GRB outburst: high accretion
rate and specific angular momentum.  This simulation provided the
first numerical confirmation of the collapsar model and showed that
with artificial, but physically reasonable, energy drivers that the
collapsar could explain the durations of LGRBs.  It is worth noting
that since this calculation, much more physically relevant models have
been run including either magnetic fields (Proga et al. 2003) or
modeling the collapse in 3-dimensions of a realistic star (Rockefeller
et al. 2006).

One way to drive the energy is through neutrino annihilation above the
black hole accretion disk.  The usual assumed geometry of a torus
accreting onto a black hole is given by a wedge, with prolific
neutrino fluxes that forms a relativistic e$^+$e$^-$ pair wind from
neutrino-antineutrino interactions.
In extreme cases for a large Shakura-Sunyaev accretion 
disk $\alpha(\approx 0.1)$ parameter and a maximally rotating black hole, efficiencies for 
$\nu \bar\nu \rightarrow $e$^+$e$^-$ could reach a few percent. 
This channel for energy release is thought to be too inefficient
to power LGRBs, though could be important for subclasses of SGRBs.
Large $a$, rapidly rotating black holes, 
can release sufficient energy to power a GRB. 
geometric beaming can be produced from energy deposition
via neutrino annihilation.  See Popham et al. (1999) for 
a review.

As far as neutrino-driven GRB engines go, the critical densities for
most likely accretion disks are $\sim 10^4$ -- $10^8$ g/cm$^3$.  For
collapsars, this corresponds to black hole masses of $\approx 10$ --
$25 M_\odot$, with delays between collapse and jet formation of 30 --
300 s.  The neutrino-driven collapsar model probably does not work
(Fryer \& M\'esz\'aros 2003). The alternatives are magnetic fields and
fast rotation (Narayan et al. 1992).  Again, see Popham et al.\ (1999)
for a review.

The magnetic field in the accretion-disk/black-hole/jet system
is probably dominated in currents flowing in the accretion 
disk. Penrose energy extraction through the Blandford-Znajek (BZ) 
process can be as powerful as the 
accretion power. 
The literature on magnetically driven jets
is extensive. 
The BZ power is insufficient except
for rapidly rotating black holes. 
Magnetically driven jets probably produce much more energy than neutrino annihilation. 
Common physics should apply to GRBs and 
AGNs with collimated relativistic
outflows, namely radio galaxies and $\gamma$-ray blazars.  

Numerical simulations of collapsar jets through stellar
envelopes of massive progenitor stars show the need for 
the envelope to have a width $\lesssim 10^{11}$ cm. Emerging
jets, possibly protected by a cocoon formed at the interface of 
the hot light relativistic fluid, can maintain relativistic speeds,
In the picture of Woosley, LLGRBs such as GRB 980425
is an off axis GRB, and XRFs could arise from angular effects in 
a mildly relativistic fireball, in contrast to 
a ``dirty," mass-loaded fireball scenario by
Dermer, B\"ottcher and Chiang (1999). ({\it Exercise:}
Tabulate similarities and differences, 
and contrast predictions of off-axis vs.\ mass-loaded 
XRF scenarios.) 
Constraints on the 
formation of collapsar GRBs include:
\begin{itemize}
\item The star must collapse to form black hole.
\item Star must lose its hydrogen envelope so that it remains compact. 
\item Jet must travel through star roughly on the GRB duration timescale.
\item Star must be rapidly rotating so disk forms around black hole.
\end{itemize}

An important mechanism 
in stellar evolution leading to GRBs is 
Roche-lobe overflow.  
When a star expands (in the giant or supergiant branch) 
in a binary, the outer layers of the star may feel 
greater gravitational attraction to the companion star, causing 
this material to ``overflow" onto that companion. A second important mechanism is
common envelope (CE) evolution.  If Roche-lobe overflow proceeds faster than the 
companion star can accrete material, the expanding star envelops the companion, 
leading to two stellar cores within a single common envelope. 
The masses of primary (most massive) and secondary (least massive) stars in a binary are $M_p$ and $M_s$.

Some pathways to GRB formation
include: (i) Massive Star which, if not affected by binary mass
transfer, would undergo core-collapse ($M_{SN}\sim 8$ -- 10 $M_\odot$)
to a black hole; (ii) He core –- He core mergers during CE phase of
binary systems with two massive stars (He core masses will also have
transition masses for neutron star and black hole formation).  The
Black Hole Mass ($M_{BH} \approx 3 M_\odot$) is the transition mass
for black hole formation, and depends on angular momentum of core. See
Fryer, Woosley, and Hartmann (1999).

In the isolated collapsar scenario, a massive single star with high
metallicity loses its hydrogen envelope via winds.  If it retains
enough mass and rotation to form a BHAD system, a GRB is
produced. Important related issues are the metallicity of the host
environment, and whether the progenitor has the required rotation.
Modeling the full collapse (first to a neutron star and then to a
black hole) is required to understand some of these trends.
Stellar collapse with nuclear equations of state are now being done in
3-dimensions (e.g. Fryer and Warren 2002) including rotation (Fryer
and Warren 2004) and finally to the late times that approach 
black hole formation (Fryer and Young 2007).  From these we can
trace fluid flow 
and isopressure 
surfaces of outward moving bubbles. ({\it Exercise:} 
Use the concepts of mixing length in convection to
write the one-dimensional stellar structure equations for a collapsing object. Discuss the equations' deficiencies.})

The combined simulation and theory understanding of stellar collapse
argues that different progenitor scenarios lead to remnant black holes
with a mass distribution possibly reflected in systems with inferred
black hole masses (Fryer \& Kalogera 2001).  Because large $a$ black
holes allow greater energy extraction per unit mass,
GRB-hosted SNe are bound to be more asymmetric than 
normal SNe to have a spun-up black hole remnant, though all SNe should exhibit asymmetries.  
Note that those objects that do form GRBs eject the star, producing a lower mass 
black hole remnant than those SNe and failed SNe that form black holes with 
no subsequent explosion (presumably due to slow rotation speeds).

One of the observed properties of GRBs is that the associated
supernovae do not show any hydrogen.  Strong stellar winds can remove
the outer hydrogen envelope, but these winds also carry away angular
momentum.  An alternative is to assume that the star is in a binary
and a common envelope phase ejects the hydrogen envelope.  This
evolution can even increase the angular momentum.  The mass and
rotation constraints are easier to satisfy in binary systems (Fryer et al. 1999).  

enough angular momentum.  An alternative, proposed by Fryer et al. (1999) 
was the merger of two stars.
Fryer \& Heger (2005) merged two helium cores and found that they could indeed 
increase the core rotation rate while removing the hydrogen and some of 
the helium envelopes.
For collapsars, this means that with magnetic fields (or with the
current prescriptions of magnetic fields), single stars and even
simple binaries have trouble working (Petrovic et al. 2005).  Yoon \& 
Langer (2005) have argued that single stars at low metallicity can 
go through extensive mixing, removing all of the hydrogen by 
burning it into helium instead of through a wind.  

But any binary may well not work.  The issue is that we not only don't see 
any hydrogen, but we don't see helium in the supernovae associated with 
GRBs.  So the models still have some work before they can fit all the 
data (see Fryer et al. 2007 for a review).

By comparison,
compact object coalescence scenarios require a binary (or multiple star) system where the 
primaries and secondaries evolve to compact objects and
merge within a Hubble time. 

Double neutron star formation for SGRBs
involve neutron star formation where the primary collapses after
expanding off the main sequence, or being triggered
by Roche lobe overflow of material onto the massive core.  
A likely scenario in close binaries involves 
CE evolution until the compact object settles to the 
center of the companion. The merger of the compact object with
a He core forms a GRB. 
Alternately, in the absence of CE evolution, 
a compact binary can form with two progenitor high-mass stars.
Accretion-induced collapse during the CE phase or during Roche
lobe overflow introduces other evolutionary pathways to GRBs.
({\it Exercise.} Report on 
a GRB progenitor pathway in more detail.)

Optimistically, a neutrino mechanism could power a GRB, 
though
would be associated with an old stellar population and 
large galactic offsets. 
Fryer et al. (1999) concluded that for collapsars, it is difficult to
make isolated stars with enough angular momentum after removing
envelope with winds.  Binaries seem necessary (e.g. He-He or
He-neutron star mergers).  Compact object mergers (NS/NS or NS/BH) are
rare, but sufficient for GRBs.  Accretion rates are high ($\gtrsim 1$
-- $10 M_\odot$/s), but short-lived ($\lesssim 200$ ms).

GRB studies intersect many exciting fields of research,
including 
\begin{enumerate}
\item Jet and fireball physics;
\item Ultra-high energy cosmic ray acceleration and neutrinos;
\item Early universe, reionization;
\item Gravitational radiation; and 
\item Mass extinctions, and geophysics. 
\end{enumerate}

Competing models include the 
\begin{enumerate}
\item Supranova and two-step (SN $\rightarrow$ NS $\rightarrow$ BH)
collapse processes for LGRBs, originally proposed by Vietri and Stella (1998,1999);
see Dermer (2008); and
\item AIC of neutron stars to black holes in binary systems as 
a model for SGRBs (MacFadyen et al.\ 2005; Dermer and Atoyan 2006).
\end{enumerate}

\section{Leptonic processes in GRBs -– prompt and afterglow emissions}
This lecture gives an overview of the elementary leptonic 
blast wave physics developed to explain GRB prompt and afterglow emissions.
The outline of this lecture 
is
\begin{enumerate}
\item Relativistic Kinematics
\item Blast-wave and Afterglow Theory
\item Relativistic Shock Hydrodynamics
\item Jetted Emission and Beaming Breaks
\item External Shocks
\item Colliding Shells 
\end{enumerate}
Reviews of GRB blast wave physics are given by M\'esza\'aros (2006),
Piran (2005), and Dermer \& Menon (2009).

To recap, we have associated
\begin{enumerate}
\item LGRBs  $\leftrightarrow$    Collapsars
\item SGRBs	 $\leftrightarrow$    Mergers of compact objects
\item LLGRBs $\leftrightarrow$   Probably a  type of collapsar, perhaps magnetar-powered
\item SGRs   $\leftrightarrow$    Highly magnetized neutron stars
\end{enumerate} 	

GRBs must be Galactic, as we now show based on discoveries 
made in the 1980s (and still valid). The typical energy flux
of a GRB is $\Phi \cong 10^{-6}\Phi_{-6}$ ergs cm$^{-2}$, with 
significant factors-of-2 variations on timescales of seconds.
Solar Maximum Mission observations of MeV photons (Matz et al.\ 1985)
showed that the optical depth for attenuation of $\gamma$ rays 
with energies $\gtrsim 1$ MeV is $\tau_{\g\g} \approx
n_\gamma\sigma_{\g\g }R \sim  n_\gamma\sigma_{\rm T }R  < 1$
for photons above threshold $\e\ep \approx 2$. The photon 
energy density $n_\g \sim u_\g/E_\g$, $u_\g \sim L_\g t_{esc}/V 
\sim 3d^2\Phi/R^2 c$, with escape time $t_{esc} \sim R/c$.
Hence 
$$\tau_{\g\g} \sim {u_\g\sigma_{\rm T} R\over E_\g}
\sim {d^2\sigma_{\rm T}\Phi\over c^2\Delta t E_\gamma} < 1\;,$$
or 
$d(10$ kpc)$ < \sqrt{\Delta t({\rm s})/\Phi_{-6}}$.
GRBs are known to be at cosmological distances. 
The flaw in this argument is the assumption that the emitting region is
at rest. (The lesson here is not ``Don't believe theorists.")
 	
The conventional fireball/blast-wave model involves
intermittent release of energy in a collimated relativistic
jet, with variations of wind parameters leading to 
internal shocks and structure in the light curves. 
External shocks are responsible for the late Beppo-SAX afterglow,
and could contribute to emission in the prompt and early
afterglow phase.

The central problem in GRB astrophysics
is to explain large apparent isotropic energies $\gtrsim 10^{54}$ 
ergs, durations $\lesssim 10^3$ s, and 
short variability time scales $\Delta t \lesssim 1$ s. The widely
accepted solution is the fireball/relativistic blast-wave model. 
The impulsive release of a huge amount of 
energy in a fireball with large entropy per baryon is described by an
\begin{enumerate}
\item Expansion phase; $\Gamma(x) \approx x/\Delta_0, x < \Gamma_0 \Delta_0$; a
\item Coasting phase; $ \Gamma(x) \cong \Gamma_0, x >\Gamma_0\Delta_0$; and a
\item Deceleration phase; $\Gamma(x) = ?$
\end{enumerate}	

\subsection{Relativistic Kinematics}

Important questions include
\begin{itemize} 
\item How to calculate $\Gamma(x)$?
\item How is $\Gamma(x)$ related to observer time $t$?
\item How to calculate internal photon number and energy density for a relativistically moving source from measured energy flux?
\end{itemize}
We consider these problems for
a simple blast-wave model with several simplifying assumptions:
a spherical, uncollimated explosion, 
a uniform surrounding medium, the blast wave approximated 
by a uniform thin shell, and particle acceleration at the 
forward shock only. Three frames of reference are considered:
the explosion (GRB, stationary, or starred) frame; the comoving fluid
(primed) frame; and the observer (unstarred) frame.
 
Using the definition of the Doppler factor
\begin{equation}
\dD = [\Gamma (1-\beta\mu)]^{-1}\;,
\label{Doppler}
\end{equation}
where  $\mu = \cos\theta$ 
and $\theta$ is the angle between the direction of the radiating
fluid and the observer,
we have from elementary considerations
\begin{equation}
dt = (1+z){d\tp\over \dD}\;,\;\;{\rm and}\;\;\e = {\dD \ep\over 1+z }\;.
\label{dtE}
\end{equation}
For well-defined pulses of radiation on a measured variability
time scale $t_{var}$,  
the comoving emission-region size scale 
\begin{equation}
r_b^\prime\lesssim  {c\dD t_{var}\over 1+z }\;.
\label{rbprime}
\end{equation}
The variability timescale 
$t_{var}$ implies an engine size scale,
with the comoving size scale a factor $\sim \dD$  larger, 
and the emission location $\sim \Gamma^2$ 
larger than values inferred for a stationary region.
Rapid variability can be obtained from an external shock
or internal shock by energizing regions within the Doppler cone, as defined by the Doppler factor of the shocked fluid.

The $\nu F_\nu$ spectrum at dimensionless photon 
energy $\e = h\nu/m_ec^2$ from a source at redshift $z$ 
and luminosity distance $d_L(z) = 10^{28}d_{28}$ cm  
is denoted by $f_\e$.
Assuming that the blob radiates isotropically in the comoving frame,
\begin{equation}
\; f_\epsilon (t) \cong 
{\dD^4 V^\prime_b\over 4\pi d_L^2}\;\epsilon^\prime j^\prime(\ep ;\tp )\;
\cong\;
 {\dD^4 \ep L^\prime(\ep ;\tp)\over 4\pi d_L^2}\;.
\label{rad,vFv}
\end{equation}
where the comoving spectral luminosity $L^\prime(\ep ;\tp) \cong
V^\prime_b j^\prime(\ep ;\tp )$, and $j^\prime(\ep ;\tp )$ is the comoving
spectral emissivity.
The spectral energy density $u^\prime ({\ep})\cong \tp_{lc} j^\prime(\ep)$.
Because $V^\prime_b = 4\pi r^{\prime 3}_b/3$,
\begin{equation}
u^\prime_{\ep } = \ep u^\prime(\ep ) = m_ec^2 \e^{\prime 2}n^\prime(\ep ) 
\cong {3d_L^2 f_\e\over \dD^4 c r_b^{\prime 2} }
\gtrsim {3d_L^2 (1+z)^2 f_\e\over c^3 \dD^6 \Delta t^2_{var} }\;.
\label{feblob5}
\end{equation}
Consequently the relation between the total internal photon energy density 
$u^\prime$ and the total measured energy flux $\Phi$ is 
\begin{equation}
u^\prime \gtrsim {3d_L^2 (1+z)^2 \Phi\over c^3 \dD^6 t_{var}^2}\;.
\label{uprimeestimate}
\end{equation}

Requiring that $\tau_{\gamma\gamma}(\e_1)< 1$, so that the emission
region is transparent to $\gamma$ rays gives, 
using a $\delta$-function result for the $\gamma\gamma$ attenuation cross section, the result
\begin{equation}
\tau_{\gamma\gamma}(\e_1)\approx {\sT\over 3}\big({2\over \ep_1}\big)
n^\prime_{ph}\big({2\over \ep_1}\big)r_b^\prime\;.
\label{tggtgg}
\end{equation}
With eq.\ (\ref{rbprime}), the requirement of optical
thinness to $\g\g$ attenuation gives for flat target photon 
SEDs in a $\nu F_\nu$ representation the result
\begin{equation}
\dD \gtrsim 200 \; \bigl[ \bigl(1+z)d_{28}\bigr ]^{1/3}
\bigl[ {\Phi_{-6} E({\rm GeV})
\over t_{var}({\rm s})}\bigr ]^{1/6}\;.
\label{dmintev}
\end{equation}
important for interpretation of Fermi Gamma ray Space Telescope (FGST) results.

\subsection{Blast Wave and Afterglow Theory}
The GRB explosion forms a radiation-dominated 
fireball with injection explosion entropy per baryon $\eta_b = L/\dot M c^2\gg 1$,
and $L$ is the wind power.
The energy of the expanding relativistic wind is transformed into
photospheric emission and the directed kinetic energy of
a hadronic shell with coasting Lorentz
factor 
\begin{equation}
\Gamma_0 = E_0/M_0 c^2 \;.
\label{M_0}
\end{equation}
Here $M_0$ is the amount of
baryonic matter mixed into the initial explosion.
For a uniform spherically 
symmetric CBM, the mass of swept-up material at
radius $x$ is $M_{sw} = 4\pi m_p n_0x^3/3$, where $n_0$ is the
proton density, assumed to be made of H.
 
The blast wave will start to undergo significant
deceleration when an amount of energy comparable to the initial baryon
energy $E_0$ in the blast wave is swept up. Looked at from the comoving
 frame, each proton from the CBM carries with it an amount of
energy $\Gamma_0 m_p c^2$ when captured by the blast wave. After capture
and isotropization, the amount of energy carried by the blast wave from
this swept-up proton is  $\Gamma_0^2 m_p c^2$ as measured in the 
stationary frame. The condition
$\Gamma_0^2 M_{sw} c^2 = E_0$ gives the deceleration radius (Rees \& M\'esz\'aros 1992;
M\'esz\'aros \& Rees 1993)
\begin{equation} 
x_d \equiv ( {3 E_0\over 4\pi\Gamma_0^2 m_pc^2 
n_0})^{1/3} \cong 2.6\times 10^{16} ({E_{52}\over \Gamma_{300}^2
n_0})^{1/3}\;\rm{cm}\;,
\label{x_d}
\end{equation}
where $E_0 = E_{52}/10^{52}$ ergs is the 
total explosion energy including rest mass energy, $\Gamma_{300}
= \Gamma_0/300$, and $n_0$ is the CBM proton density in units of cm$^{-3}$.

Differential time elements in the stationary (starred), comoving (primed), and 
observer (unscripted) reference frames satisfy the relations 
$$dx = \beta c dt_* = \beta \Gamma cdt^\prime = \beta c\;
{dt\over (1+z)(1 -\beta \mu)}\;,$$ 
where the last expression is obtained by
noting that $dt/(1+z) = dt^\prime (1-\beta\mu)$, and $\theta =
\arccos\mu$ is the angle between the direction of outflow and the
observer. Hence
\begin{equation}
dt = {(1+z)\over c}  dx \;(\beta^{-1} - 
\mu) \cong {(1+z) dx\over \Gamma^2 c}\;.
\label{t}
\end{equation}
The last expression applies to relativistic flows ($\Gamma \gg
1$) observed on-axis, assuming that the average emitting region is at
$\mu \cong \beta$.

The deceleration time 
as measured by an observer is therefore
\begin{equation}
t_d \equiv (1+z) {x_d\over  \Gamma_0^2 c}\;
\cong {9.6\; (1+z)\over \beta_0}\;({E_{52}\over 
\Gamma_{300}^8 n_0})^{1/3}\;\rm{s}\;.
\label{t_d}
\end{equation}

\subsection{Blast-Wave Equation of Motion}

The equation describing the speed of the relativistic blast wave, 
which changes as 
a consequence of the blast wave sweeping up material from the surrounding medium, 
is for an adiabatic blast wave, 
$$ \Gamma[M_0 + \Gamma m(x)] \cong \Gamma[M_0 + kx^3(\Gamma -1 )]\cong
~ const \;,$$ 
where $m(x)$ is the swept-up mass. 
For $\Gamma(x) \propto x^{-3/2}$, $t \cong c^{-1} \int dx\;\Gamma^{-2} 
\propto \int dx\; x^3$, so $x(t) \propto t^{1/4}$ and $\Gamma(t) \propto
t^{-3/8}$.
If the blast wave is partially or highly radiative, different 
behaviors follow. [{\it Exercise:} Derive the power-law behavior for adiabatic 
blast waves decelerating in an external medium with 
radial wind density profiles.]

The kinetic energy swept into the comoving fluid frame per unit
proper time at the forward shock is given by
\begin{equation}
{dE^\prime\over dt^\prime}|_{FS} = A(x) n_0 m_pc^2
(\beta c)\Gamma(\Gamma-1)\;\propto \Gamma^2~~{\rm for}~~
\Gamma \gg 1\;.
\label{dEdt}
\end{equation} 
where the area $A(x) = 4\pi x^2$ for an 
isotropic blast wave. The factor $\Gamma$ represents the increase
of external medium density due to length contraction, the factor
$(\Gamma -1)m_p$ is the kinetic energy of the swept-up
particles, and the factor $\beta c$ is proportional to the rate at 
which the particle energy is swept into by the blast wave. This process provides internal energy
available to be dissipated in the blast wave. 
The original treatment of  adiabatic and radiative relativistic blast waves
using a fluid dynamical approach was given by
Blandford and McKee (1976).

A fraction $e_e$ of the forward-shock power
is assumed to be transferred to the electrons, so that
\begin{equation}
L_e^\prime = \e_e {dE^\prime\over dt^\prime} \;.
\label{Le}
\end{equation}
If all the swept-up electrons are accelerated, then joint normalization to power and number gives
\begin{equation}
\gamma_{min} \cong \e_e({p-2\over p-1})({m_p\over m_e})(\Gamma-1)\;\cong \e_e({p-2\over p-1})({m_p\over m_e})\Gamma~\;,\: {\rm for}~~
\Gamma\gg 1 
\label{gminap}
\end{equation}
and $2 < p < 3$.
The magnetic-field energy
density $u_B = B^2/8\pi$ is assumed to be a  fixed fraction $\e_B$ of the downstream energy density of
the shocked fluid. Thus
${B^2/ 8\pi} \cong 4\e_B n_0 m_pc^2\Gamma^2\;.$ 
A break is formed in the electron spectrum at cooling electron Lorentz factor $\gamma_c$,
which is found by balancing the synchrotron loss time scale $t_{syn}^\prime$ with the
adiabatic expansion time $t^\prime_{adi} \cong x/\Gamma c\cong \Gamma t \cong t_{syn}^\prime
\cong (4c\sigma_{\rm T} B^2 \gamma_c/24\pi m_e c^2)^{-1}$, giving
\begin{equation}
\gamma_c \cong {3m_e\over 16\e_B n_0 m_p c\sigma_T\Gamma^3 t}\;.
\label{gamma_c}
\end{equation} For an adiabatic blast wave, $\Gamma\propto t^{-3/8}$, so that 
$\gamma_{min} \propto t^{-3/8}$ and $\gamma_c \propto t^{1/8}$.

The observed $\nu F_\nu$ synchrotron spectrum from a GRB depends on the 
geometry of the outflow. If $L^\prime_{syn}(\ep )=\ep (dN^\prime/d\ep dt^\prime)$ 
is the spectral luminosity in the comoving frame, then $\epsilon^\prime L^\prime _{syn}(\ep)
\cong {1\over 2} u_B c\sigma_{\rm T} \gamma^3 N_e(\gamma)$, with $\gamma = \sqrt{\ep/\e_B}$ and  
$\e_B = B/B_{cr} = B/(4.41\times 10^{13}$ G). 
For a spherical blast-wave geometry, the spectral power is amplified by two powers of 
the Doppler factor $\delta$ for the transformed energy and time.
The $\nu F_\nu$ synchrotron spectrum is therefore 
\begin{equation}
f_\epsilon^{syn}\cong {2\Gamma^2\over 4\pi d_L^2}\;(u_Bc\sigma_{\rm T})\; 
\gamma^3 N^\prime_e(\gamma)\;,\;\gamma \cong \sqrt{{(1+z)\e\over 2\Gamma\e_B}}\;.
\label{fe}
\end{equation}

For a power-law injection spectrum, the cooling comoving nonthermal
electron spectrum can be approximated by 
\begin{equation}
N^\prime_e(\gamma) \cong  {N_e^0\;\gamma_0^{s-1}\over s-1}
\cases{\gamma^{-s}\; ,&  $\gamma_0 \lesssim \gamma \lesssim \gamma_1$ \cr\cr
	\gamma_1^{p+1-s}\gamma^{-(p+1)} \; , & $\gamma_1 \lesssim \gamma \lesssim \gamma_2.$  \cr}
\label{Ne}
\end{equation} 
In the slow cooling regime, $s = p$, $\gamma_0 = \gamma_{min}$ and $\gamma_1 = \gamma_c$, whereas
in the strong cooling regime,  $s = 2$, $\gamma_0 = \gamma_c$ and $\gamma_1 = \gamma_{min}$. 
This method for deriving the 
behaviors of the breaks in the electron and synchrotron 
spectrum for the simple synchrotron blast-wave model
were originally given by Sari, Piran, and Narayan (1998). 
[{\it Exercise.} Derive the temporal
and spectral behaviors for the elementary blast-wave model
consisting of an impulsive adiabatic blast wave with a strong
forward shock sweeping a uniform surrounding medium.] 

The synchrotron shock blast-wave model has been used to fit
afterglow data and 
deduce microphysical and environmental parameters. 
Detailed leptonic models include a synchrotron self-Compton (SSC) component, 
which is highly sensitive to the baryon-loading parameter $\Gamma_0$. 

\subsection{Beaming Breaks and Jets}

An observer will receive most emission from those portions of a GRB
blast wave that are within the Doppler angle $\theta_{\rm D} \sim 1/\Gamma$ to the direction
to the observer.  As the blast wave decelerates by sweeping up
material from the external medium, a break in the light curve will occur when the
jet opening half-angle $\theta_j < 1/\Gamma$. This is due to a change from a spherical blast wave
geometry to a geometry defined by a
localized emission region).
Assuming that the blast wave decelerates adiabatically in a uniform
surrounding medium, the condition $\theta_j \cong 1/\Gamma =
\Gamma_0^{-1}(x_{br}/x_d)^{3/2} = \Gamma_0^{-1}(t_{br}/t_d)^{3/8}$
implies
\begin{equation}
t_{br} \approx 45(1+z)\;({E_{52}\over n_0})^{1/3}\theta_j^{8/3}\;{\rm days}\;,
\label{tbr}
\end{equation}
from which the jet angle 
\begin{equation}
\theta_j \approx 0.1 [{t_{br}({\rm d})\over 1+z}]^{3/8}\;
({n_0\over E_{52}})^{1/8}
\label{thetaj}
\end{equation}
can be derived. Note that the beaming angle is only
weakly dependent on $n_0$ and $E_0$.

Numerical models show X-ray beaming breaks hidden by the effects
of the SSC component. 
The important discovery (Frail et al.\ 2001) of a clustering, 
beaming-corrected energy for LGRBs opens the possibility to perform 
cosmological studies with GRB data.

\subsection{Relativistic Shock Physics}

The structure of a shock is determined
by continuity of the particle number, energy
and momentum  fluxes across the shock
front.  The
 pressure $p = (\hat \gamma - 1)e_{ke}$, where $e_{ke}$ is 
the kinetic energy density of the fluid, and $\hat \gamma$ corresponds to
the ratio of specific heats.  For a nonrelativistic monatomic ideal
gas, $\hat \gamma = 5/3$, whereas $\hat \gamma$ = 4/3 for a
relativistic gas. For strong shocks, $ n^\prime \cong 4\Gamma n_0$.

The equality of kinetic-energy
densities at the contact discontinuity implies, for fluids made primarily
of proton-electron plasma, that
\begin{equation}
{e_{ke}\over m_pc^2}\cong n_f(\Gamma - 1)  
\cong n_r (\bar\Gamma - 1 )\;\cong \; 4 n_0\Gamma^2\;
\cong \; 4n(x)(\bar\Gamma^2 -\bar\Gamma),
\end{equation}
where the reverse shock Lorentz factor is $\bar \Gamma  = \sqrt{1-\bar\beta^2}$. 
The relativistic shock jump conditions for an isotropic explosion in a
uniform surrounding medium imply (Sari \& Piran 1995)
\begin{equation}
{n(x)\over n_0} \equiv F = 
{E_0\over 4\pi x^2 \Gamma_0^2 n_0 m_pc^2
 \Delta}\cong{\el^3\over x^2 \Gamma_0^2 \Delta}\;= {\Gamma^2\over
 \bar\Gamma^2-\bar\Gamma}\rightarrow \cases{2\Gamma^2/\bar\beta^2\; ,&
 ${\rm ~~NRS}$ \cr\cr \Gamma^2/\bar\Gamma^2\; ,& ${\rm ~~RRS}$
 \cr}\;.
\label{n(x)/n}
\end{equation}
The relations between the forward shock (FS) and reverse shock (RS) Lorentz factors 
can be derived in the limit of a nonrelativistic reverse shock (NRS) and strong
forward shock, and in the limit of a relativistic reverse shock (RRS) and 
relativistic forward shock. 
The RS power is $dE^\prime/dt^\prime|_{RS} = A(x)
n(x)m_pc^3 \bar\beta(\bar\Gamma^2 -\bar\Gamma)$. 
With the shock jump condition, 
one finds that 
$(dE^\prime/dt^\prime|_{RS})/(dE^\prime/dt^\prime|_{FS}) = \bar\beta$,
so that roughly equal power is dissipated as internal energy in the
forward and reverse shock during the RRS phase.

\subsection{External Shock Model}

This is sufficient blast-wave 
physics that evolving forward and reverse shock emissions can be 
calculated for comparison with data. 
Synthetic light curves
in the external-shock model can be derived and used to make predictions
for the generic behavior of GRBs with smooth light curves.
Spikiness of the light curve could originate from inhomogeneities
in the external medium, before the blast wave has entered the 
adiabatic deceleration regime. 
Consider a blast wave 
intercepting a cloud with size 
$r \ll R/\Gamma_0$ that is located at an angle $\theta$ 
with respect to the line of sight to the observer. The duration of the received pulse of radiation depends on the light travel-time delays from different portions of the blast wave as it interacts with the cloud. Photons emitted when the blast wave passes through the near and far sides of the cloud are received over a {\it radial} timescale 
\begin{equation}
t_r = {2r\over \beta_0\Gamma_0\dD c}\cong {r\over \Gamma_0^2 c}\;.
\label{tr}
\end{equation}
Photons emitted from points defining the greatest angular extent of the cloud are received over an {\it angular} timescale
\begin{equation}
t_{ang} \cong {r\theta\over c}\;.
\label{tang}
\end{equation}
Note that if $r \rightarrow R/\Gamma_0$ and $\theta \rightarrow 1/\Gamma_0$, then $t_{ang} \rightarrow R/\Gamma_0^2 c$, as expected. When $\theta \approx 1/\Gamma_0$, $t_{ang} \approx \Gamma_0 t_r\gg t_r$. Except for those few clouds with $\theta \lesssim 1/\Gamma_0^2$ lying almost exactly along the line of sight to the observer, $t_{ang} \gg t_r$. 
Highly variable light curves with reasonable ($\gtrsim 10$\%) efficiency
can be produced in an external shock model (Dermer \& Mitman 1999).

In the external shock model, 
 a single relativistic wave of particles interacts 
with inhomogeneities in the surrounding medium to 
accelerate particles that radiate the prompt $\gamma$ rays.
A central requirement for strong radiative efficiency in an 
external shock model for the prompt phase is that a strong 
forward shock is formed; otherwise the Lorentz factor $\Gamma\ll \Gamma_0$
and the radiation is strongly debeamed. A strong forward shock is formed
when the comoving shell density 
$n(x)\gg \Gamma_0^2 n_0$.

\subsection{Internal Shock/Colliding Shell Model}

In the internal shock model, an active central engine 
eject waves of relativistic plasma that overtake and collide 
to form shocks. The shocks accelerate nonthermal particles 
that radiate high-energy photons.
The relative Lorentz factor of the two shells
with Lorentz factors $\Gamma_1$ and $\Gamma_2$, with $\Gamma_2 = \zeta \Gamma_1$, 
$\zeta \geq 1$,  
is 
\begin{equation}
\Gamma_{rel} = \Gamma_1\Gamma_{2}(1-\beta_1\beta_{2}) 
\stackrel{\Gamma_1\Gamma_{2} \gg 1}{\;\;\;\;\longrightarrow\;\;\;\;\;}{1\over 2}
\;\big(\zeta +\zeta^{-1}\big)\;.\label{Gammarel}
\end{equation}
For mildly relativistic internal shocks with a range of 
relative Lorentz factors $1\lesssim \Gamma_{rel}, \zeta \lesssim 10$, 
 the Lorentz factor $\hat \Gamma$ of the shocked fluid with adiabatic 
index $\hat\gamma = 5/3$ in the explosion 
frame is
$$ \Gamma_{sf} \cong 2\Gamma_1\;{\Gamma_{rel}F^{1/4}\over 
\sqrt{2\Gamma_{rel} - F^{1/2}}}\;\cong \sqrt{2\Gamma_{rel}}\Gamma_1\;\cong \sqrt{\Gamma_1\Gamma_2}\;,$$
where $F = n_2/n_1$ is the ratio of proper frame densities of shell 2 to 
shell 1 when they intercept each other (eq.\ [\ref{n(x)/n}]), 
and the final expression assumes
that $n_2\approx n_1$ and $\Gamma_{rel}\gg 1$.
The proper shocked fluid number and energy densities are 
$$ n_{sf} = (4 \Gamma_{rel} +3)n_1\;\approx 4\Gamma_{rel} n_1\;,\;{\rm~and~}\;
e_{sf} = \Gamma_{rel} n_{sf}\;m_pc^2\;.$$

Now consider the elastic collisions of shell 2 with mass $m_2$ intercepting 
shell 1 with mass $m_1$. In an elastic collision, the Lorentz factor of
the merged shell is 
\begin{equation}
\Gamma_m\;\cong \; \sqrt{ {m_1\Gamma_1 + m_2\Gamma_2\over m_2/\Gamma_2 + m_1/\Gamma_1}}\;.
\label{Gammamerge}
\end{equation}
The efficiency to convert the directed kinetic energy 
of the shells into internal energy is 
\begin{equation}
\eta\; =  \; 1 -  {(m_1+m_2)\Gamma_m\over m_1\Gamma_1 + m_2 \Gamma_2}\;.
\label{efficiency}
\end{equation}
The efficiency is greatest when the shells have comparable mass and
$\Gamma_2\gg\Gamma_1$; otherwise $\eta\sim few$ \%. 
When the contrast between the 
$\Gamma$ factors of the shells is large, $\eta \sim 10$ -- 20\% is possible. 

The rapid X-ray decline can be explained by high-latitude
emission after the central engine has been turned off.
Writing the flux density $F_\nu \propto \nu^\alpha t^\beta$
gives the curvature relation, which assumes that the spectral shape of
the radiated flux is constant within the Doppler cone. The
curvature relation is $\alpha = 2-\beta$.

In the standard model for LGRBs with relativistic winds and colliding
shells, X-ray flares are made when the GRB engine is restarted. 
Long-lasting GRB central engines can also involve 
continual injection scenarios with pulsars. 
The generic long-duration GRB
light curve at $\sim 1$ keV, from Swift observations (Zhang et al.\ 2006;
Nousek et al.\ 2006), can 
be divided into a prompt phase (0), a decay phase (I), a plateau phase (II), 
 an afterglow phase (III), and a jet break phase (IV), in addition 
to X-ray flares (V). 
Kinematic shapes of 
GRB pulses can be calculated for 
illuminated shells 
when the thickness and duration of illumination 
of the shell are varied. 

\subsection{Leptonic GRB Physics: Summary}

The success of the fireball/relativistic blast wave model 
arises from its ability to explain
\begin{enumerate}
\item Large energy releases in short times;
\item Escape of $\gamma$ rays;
\item Afterglows at various wavelengths from radio through $\g$ rays.
\end{enumerate}
The physics is 
widely applicable to many nonthermal systems, including
blazars, microquasars, pulsar winds,\dots

This hardly exhausts leptonic blast-wave physics. Other interesting
physics involves thermal photospheres, the ``line of death" and synchrotron 
jitter radiation, and the origin of the Amati and Ghirlanda relations. 

\section{Hadronic Processes and Cosmic Rays in GRBs}

Acceleration of ultra-relativistic 
protons and ions is favored in a blast wave
physics scenario at least as much energetic ions, 
insofar as $(1-\e_e)$ of the power in the dissipation
region emerges in the form of  
magnetic field energy or ions and, if ions, 
with energy $\gtrsim \Gamma_0^2$ GeV.  
The outline for this lecture is
\begin{enumerate}
\item Ultra-high Energy Cosmic Rays (UHECRs)
\item Photohadronic Processes, Energetics, and Power
\item Hadronic Blast Wave Theory
\item Cosmic Rays from GRBs
\item Neutrinos from GRBs
\item GRBs in the Milky Way 
\end{enumerate}

\subsection{Ultra-High Energy Cosmic Rays}

The year 1912 is a landmark date in space
science when the cosmic radiation, a ``penetrating
radiation from above," was discovered 
by Victor Hess by flying electroscopes on a balloon.
The cosmic-ray energy density at GeV/nucleon
energies is $u_{CR} \approx 10^{-12}$ ergs/cm$^{3}$, 
with the total cosmic-ray kinetic energy 
density modulated by the outflowing Solar wind 
on 22-yr Solar cycle.  The knee of the spectrum 
is at $\approx 3$ PeV, the second knee is at particle
energy $E\approx 10^{17.4}$ eV, the ankle (or dip) 
at $E\approx 10^{18.6}$ eV, and the GZK cutoff is at 
$E_{\rm GZK} \approx 10^{19.5}$ eV. The value of 
$u_{CR}\approx 10^{-21}$ ergs/cm$^{3}$  at 
$E\gtrsim E_{\rm GZK}$. 

The Pierre Auger Observatory (PAO), located in the 
Mendoza Province in Argentina at 
$\approx 36^\circ$ S latitude
determines the arrival directions and energies of UHECRs 
using a hybrid technique consisting of 
four telescope arrays to measure Ni air fluorescence 
and 1600 surface detectors spaced 1.5 km apart to measure muons 
formed in cosmic-ray induced showers. 
Event reconstruction using the hybrid technique 
gives arrival directions better than $1^\circ$, 
and energy uncertainties better than $\approx 20$\%. 
Two important discoveries were made in 2007, namely
\begin{itemize}
\item GZK cutoff with the HiRes Observatory (Abbasi et al.\ 2008) and the PAO (2008).
\item Clustering of arrival directions toward AGN in the supergalactic plane, with the PAO (2007).
\item Interesting though ambiguous results on composition were also announced
by the PAO in 2007 (Unger et al.\ 2007).
\end{itemize}

The GZK cutoff, now seen clearly with HiRes and the PAO, 
contrary to earlier AGASA results, is a consequence of exponentially increasing energy 
losses of UHECR protons due to photopion-producing reactions with photons
of the CMBR. For UHECR Fe, strong photo-dissociation though, primarily, one
and two-nucleon losses in giant dipole resonance reactions, also produce a
GZK cutoff. 
There is no strong evidence for a lighter composition 
at $E \gtrsim 10^{19}$ eV inferred from data taken with the PAO to 
energies $E \lesssim 4\times 10^{19}$ eV, contrary to pre-Auger suggestions
(Watson 2006).
Extrapolation of particle physics uncertainties 
to large values of total CM energy make deductions about composition
from shower data uncertain.
 
A major result announced by the PAO collaboration in 2007 was
the correlation of arrival directions of $E\gtrsim 60$ EeV  UHECRs
with nearby, $d \lesssim 75$ -- 100 Mpc,  AGNs in the V{\'e}ron-Cetty
\& V{\'e}ron (2006) catalog. A marked excess in the direction of Cen A 
has generated much interest in the possibility that Cen A is 
an UHECR source. 
Both radio-loud AGN and GRBs remain
plausible candidates, but the absence of strong radio galaxies 
within $\approx 100$ Mpc of a number of UHECR arrival directions is 
puzzling for a radio-galaxy origin of the UHECRs.

\subsection{Photohadronic Processes, Energetics, and Power}
 
Like filling a bathtub, filling the Galaxy with cosmic
rays, or the universe with UHECRs, is a balance between injection
and escape (loss). Symbolically,
\begin{equation}
u_{CR} (E) \simeq t_{loss}(E) \dot\varepsilon_{CR}(E)\;.
\label{uCR}
\end{equation}
The emissivity $\dot\varepsilon_{CR}(E)$ is source dependent, and 
expected to favor strong nonthermal sources. Relevant 
photohadronic proceesses are photomeson and
photopair production, and photodisintegration
for ions. 
The photomeson or photopion cross section resembles
a threshold  step-function due to the onset of various resonant
and nonresonant channels above threshold. Photopion
losses due to interactions with 
photons of the extragalactic background light (EBL), importantly
consisting of the CMBR,  and Bethe-Heitler ($N\gamma \rightarrow N$e$^+$e$^-$)
photopair losses on nucleon $N$. In addition, one must consider 
ion-synchrotron and universal expansion losses. ({\it Exercise:}
Derive the ion-synchrotron energy loss rate in a tangled magnetic
field, and the maximum synchrotron photon energy and particle energy
in Fermi acceleration scenarios.) For photopion losses, we use
a step-function approximation, from which the energy loss mean-free-path
for particles of energy $E$ can be derived. 

The UHECR emissivity (or luminosity density), from eq.\ (\ref{uCR}), is 
$\dot\varepsilon_{CR}(E) \simeq u_{CR} (E)/ t_{loss}(E)$. 
Using PAO measurements and results of energy-loss mean-free-paths, 
then $\dot\varepsilon_{CR}(E)\simeq
10^{44}$ ergs/Mpc$^3$-yr for $E\gtrsim 10^{18}$ eV.
This is within an order of magnitude of the 
electromagnetic emissivity of GRBs, as noted in 1995 
by Vietri and Waxman. If the baryon-loading factor $f_b\gg 1$, 
as implied when $\e_e\ll 1$, then LGRBs could power the UHECRs; 
SGRBs do not seem to have the required emissivity.

LLGRBs, which take place at low redshifts (GRB 980425/SN 1998bw was at 
$d\approx 40$ Mpc), can also, in principle have the requisite emissivity.
Estimates show emissivities of the outflowing kinetic energy in LLGRBs
of $\approx 250\times 10^{44}$ ergs/Mpc$^3$-yr (Wang et al.\ 2007); however only a small fraction
of this luminosity density can be expected to emerge in the form of 
hard X-rays and soft $\gamma$ rays.

\subsection{Hadronic Blast Wave Theory}

The elements of blast-wave theory applied to leptons is, with
appropriate changes, directly applied to hadrons. In the simplest
model, cosmic-ray protons are injected downstream of the shock
with spectrum
\begin{equation}
\dot N^\prime (\gp ) \propto \g^{\prime -p}H(\gp; \gp_{min}, \gp_{max})\;,
\label{Nprime}
\end{equation}
normalized to the number of swept-up protons $N_0 = 4\pi x^3/3$, 
the swept-up power, and the magnetic field $B$ defined in terms 
of shocked fluid energy density (Lecture 3). The minimum comoving
proton Lorentz factor $\gp_{min} \approx \Gamma(x)$, and the maximum
is set by equating size scale with Larmor radius. 
 For equipartition magnetic fields, cosmic-ray protons
reach ultra-high, $\gtrsim $ EeV energies after freely escaping from
the blast wave into interstellar space. ({\it Exercise:} work
out these relations.)

The emission spectrum from a GRB includes
in addition to a leptonic component a photohadronic 
component.  The three dominant collision processes
are (i) secondary nuclear production; (ii) photomeson
production; and (iii) photopair production. Energetics
arguments favor photohadronic processes over the
 nuclear collision processes.
 Proton and ion synchrotron radiation 
must also be considered. 
The cascade $\gamma$-ray spectrum is initiated by
 decay of secondary
mesons, attenuation of high-energy photons, 
and synchrotron radiation. 
Hadronic emission 
decays more slowly than leptonic emissions in 
standard blast-wave model calculations (B\"ottcher \& Dermer 
1998). ({\it Problem:} 
Analytically examine correlations in variability behavior
for proton synchrotron $\gamma$ rays and leptonic
synchrotron emission. Do the same for photohadronic
processes.)

\subsection{UHECRs from GRBs}

The astrophysics to calculate the UHECR energy spectrum
measured here at Earth can be extensive (Berezinskii \& Grigor'eva 1988). 
Assuming a homogeneous source type that explodes and releases 
the same UHECR spectrum into intergalactic space throughout
cosmic time enormously simplifies the problem. 
The 
star formation rate factor for GRB progenitors could follow
the classical SFR (Hopkins \& Beacom 2006), 
or have a different dependence due, e.g.,  metallicity effects. 
For very active SFRs at $z \approx 1$
compared to now, the pair-production trough becomes more pronounced
to explain the ankle feature through photopair losses. 
The model of Wick, Dermer, and Atoyan (2003) for the
UHECR spectrum from GRB sources predicts a very sharp GZK 
cutoff but requires a large baryon load, 
$f_b \approx 30$ -- 100. (Note 
that the lower normalization of Auger vs.\ HiRes brings the 
value of $f_b$ down.)

Different choices for the GRB star formation rate, which 
normalizes the luminosity density (emissivity) at different $z$, 
can be normalized to various SFR factors. Very active
star formation rates 
can be ruled out by comparing Swift and pre-Swift
statistical GRB data (Le \& Dermer 2007). 
The UHECR spectrum is also quite
sensitive to the maximum energy of accelerated ions or protons.

\subsection{Neutrinos from GRBs}

The opening of the high-energy ($>$ TeV) neutrino window is anxiously
awaited, when statistically significant detection of a cosmic point or extended
source of high-energy neutrinos 
with a deep ice-based or water-based Cherenkov light sensitive
detectors happens. Charged-current interactions of a neutrino with a nucleon
induce muon, electron, or tau production(Gaisser et al.\ 1995). These particles cascade and 
shower to make Cherenkov light. 
Detector optical modules (DOMs) in the 
IceCube detector at the South Pole detect upward-going tracks to 
screen out the intense background from downward-going showers induced
by cosmic-ray muons. 

Significant extraction of UHECR energy via photohadronic
processes can be made in a collapsar-model GRB, with a significant
fraction going into neutrinos. 
The efficiency for 
neutrino production depends importantly on the baryon loading and 
Doppler factor. A spectrum with index $-2$ minimizes energy requirements
and gives detectable neutrino production for the large GRB baryon loading
required to fit the UHECR spectrum. 
Interesting anti-correlations
between neutrino and $\gamma$-ray brightness arise because of the
underlying physics.  
Photomeson cascade calculations 
consist of many generations of Compton-upscattered and lepton 
synchrotron radiation. 
A $\gamma$-ray spectral model of 
a GRB must include, at least, a photohadronically induced $\gamma$-ray 
component from UHECRs in GRB blast waves, in addition to the lepton
synchrotron and SSC radiation. The high-energy neutrinos are formed
between $\approx 100$ TeV and 100 PeV.

Photomeson production in a GRB blast wave, for parameters 
used to fit the UHECR spectrum, leads inevitably to a neutral 
beam of neutrons, $\gamma$ rays, and neutrinos.  Many interesting
implications of the neutral beam model (Atoyan \& Dermer 2003)
follow, first being detectability of GRBs in high-energy 
neutrinos for reasonable collapsar model GRB parameters, though 
with Doppler factors $\lesssim 100$. 
Subsequent photopion interactions
from UHECR neutrons or neutron-decay UHECR protons induce 
beam of $\gamma$ rays and leptons that cascade, making hyper-relativistic, 
highly polarized synchrotron radiation.
A classical LGRB
model for UHECRs allows fairly definite predictions to be made
involving only a few parameters
for the cosmogenic GZK neutrino spectrum. This gives the ``guaranteed" 
UHECR $\nu$ spectrum for a given astrophysical/cosmological model of
UHECR origin.

\subsection{GRBs in the Galaxy}

\subsubsection{Cosmic Rays from GRBs in the Galaxy.}

Depending on the beaming factor $\varphi_{bm}$, the
rate of GRBs in Milky Way is estimated to be $\sim 1$ per (0.1 -- 1) Myrs.
The rate of GRBs producing astrobiological effects at Earth is $\sim 1/$Gyr.

For a beaming factor $\sim 1$/50 -- $ 1$/500 (Frail et al.\ 2001)
the	mean $\gamma$-ray energy in X/$\gamma$-rays is $\approx 5\times 10^{50}$ ergs.
The likelihood of a recent ($\lesssim $ Myr) GRB in our Galaxy is 
scaled from BATSE rate of $\approx  2$ GRB/day. From these estimates,   
$\approx 0.3$ -- 1\% of SNe collapse into black holes, implying   
$\sim 1$ GRB  every $\sim 3$ -- 100 kyrs in the Galaxy 
The expected number of recent GRBs within $r($kpc) of Earth over a period
of duration $t$ is, roughly,
\begin{equation}
\langle N_{GRB}\rangle\simeq r^2({\rm kpc})\, t({\rm Myr})\;.
\label{NGRB}
\end{equation}
({\it Exercise:} Verify, correct, or improve these estimates.)
 
Upon injection by a GRB, high-energy cosmic rays diffuse
throughout the disk into the halo of the Milky Way. 
For isotropic turbulence and a diffusion
coefficient $D(E)$, a particle cloud diffuses according to the relation
\begin{equation}
 N(E,r,t) \propto r_{diff}^{-3} \exp{(-r^2/r^2_{diff})}\;\;,\;\;r_{diff}(E)  = \sqrt{2D(E) t}.
\label{NErt}
\end{equation}
A two-component turbulence spectrum, corresponding to turbulent energy injection at the pc and 
100 pc scales following Kolmogorov and Kraichnan behaviors, was used to 
model (Wick et al.\ 2004) 
propagation of cosmic rays and ions between $\approx 100$ TeV and $\approx 10^{17}$ eV. 
The diffusion coefficient is limited by the relation $r_{diff}(E)/t < c$ and, 
in general, is anisotropic.

In its simplest form, such a model is sufficiently 
quantitative to fit the energy spectra of UHECR ions, as measured with the 
Karlsruhe observatory, KASCADE (Antoni et al.\ 2004). The injection spectrum of the ion component
is fixed, with composition varied to improve the fit. The propagation characteristics
depends of the rigidity coefficient, essentially energy per charge at these energies.
A crucial issue is anisotropy. The combined Galactic GRB model
for cosmic rays through the knee, combined with the 
extragalactic/cosmological UHECRs make a complete model 
for high-energy cosmic rays.
Propagation of high-energy cosmic rays in the 
Galaxy exhibit surprising but easily understood effects.

\subsubsection{Astrobiological Effects from GRBs in the Galaxy.}

The rate of intense events from GRBs in the Galaxy can be estimated
from the previous results (Dermer \& Holmes 2005). 
Define the bolometric photon fluence $\varphi
=\Sc\varphi_{\odot}$ with reference to the Solar energy fluence
$\varphi_{\odot} = 1.4\times 10^{6}\Sc$ ergs cm$^{-2}$ received at
Earth in one second.  Significant effects on atmospheric chemistry
through formation of nitrous oxide compounds and depletion of the
ozone layer is found when $\Sc \gtrsim 10^2$ -- $10^3$
because of the very hard incident
radiation spectrum of GRBs that is reprocessed into 
biologically effective 200 -- 320 nm UV radiation.

Using the standard energy reservoir for LGRBs to 
establish apparent energy release for a jet
with opening half-angle $\theta_j$,
one finds that the maximum sampling distance $R_s$
of a GRB with apparent isotropic $\gamma$-ray energy release
$E_{\gamma,iso}$ to be detected at the fluence level $\varphi >
\varphi_{th}=\Sc\varphi_{\odot}$ is
\begin{equation}
R_{s} = \sqrt{{E_{\gamma,iso}\over 4\pi \varphi_{th}}}\cong
{1.1 {\rm~kpc}\over (\theta_j/0.1)}\sqrt{{\Ec_{51}\over (\Sc/10^3)}}\;.
\label{rsamp}
\end{equation}

If one GRB occurs every
$10^5 t_5$ years in the Milky Way, the rate
of biologically significant events is
\begin{equation}
\dot N(>\Sc ) \simeq {0.3\over R_{15}^2} \;
{\Ec_{51}\over (\Sc/10^3) t_5}\; {\rm Gyr}^{-1}\;.
\label{nsc_1}
\end{equation}
Thus a GRB at a distance
$\approx 1$ kpc with $\Sc \gg 10^2$ takes place about once every
Gyr, and more frequently if $t_5\cong 0.1$.

A GRB pointed towards Earth produced a lethal flux of high-energy photons and 
muons that destroyed the ozone layer, killed plankton, and led to trilobite 
extinction in the Ordovician Epoch (Melott et al. 2004). 
Geological evidence points toward two pulses: 
a prompt extinction and an extended ice age.
A muon dose and intense flux of ionizing photons 
from a GRB could have produced the prompt extinction.  
Delayed cosmic rays could have produced the later ice age.
({\it Exercise:} Calculate muon dose at Earth's surface
from UHECR neutron impacts on the upper atmosphere made by a 
GRB in the Galaxy. Describe effects.)

The issue of metallicity-dependence of LGRB progenitors
introduces a large additional uncertainty into these estimates
(Stanek et al. 2006).
Local LLGRBs are found in low-metallicity host galaxies. 
This may also be the case for LGRBs. Recent work has also 
considered the role of LLGRBs as sources of UHECRs.

A complete model for cosmic rays developed by Wick, Dermer, and Atoyan (2004) has
\begin{enumerate}
\item Cosmic Rays below $\lesssim 10^{14}$ eV from SNe that collapse to neutron stars
\item Cosmic Rays above $\gtrsim 10^{14}$ eV from SNe that collapse to black holes
\item CRs between knee and ankle/second knee from GRBs in Galaxy
\end{enumerate}
The highest energy cosmic rays originate from outside our Galaxy, because their
Larmor radii exceed the size scale of the Galaxy.
({\it Problem:} Derive the transition energy from galactic to cosmological dominance
of cosmic rays for a realistic galactic magnetic field model. 
Fit to data using simplifying assumptions.)

\subsection{Hadronic GRBs: Summary}

The acceleration of ultrarelativistic hadrons in GRB blast waves
introduces many new 
aspects to the GRB problem. Specializing to LGRBs and 
UHECR protons, we find that
\begin{itemize}
\item LGRBs are a viable source of UHECRs; 
\item Hard $\gamma$-ray emission components are formed by hadronic 
cascade radiation inside GRB blast wave; 
\item	A second emission component is formed by outflowing high-energy neutral
 beam of neutrons, $\gamma$-rays and neutrinos; 
\item GRBs can be detectable high-energy neutrino sources, which would confirm UHECR acceleration;
\item Cosmic rays can be accelerated by GRBs in the Galaxy; and 
\item  GRBs could be responsible for ionizing or extinction events on Earth.
\end{itemize}
A Northern hemisphere Auger, like its Argentine counterpart, 
will produce the first full-sky map of the universe not in the 
electromagnetic window, but in the particle window.

\section{GRB and $\g$-ray studies with Fermi/GLAST}

Each major advance in GRB science is a result of 
new instrumentation. The next GRB epoch has already began
with the launch of GLAST on June 11, 2008, renamed
the Fermi Gamma-ray Space Telescope (FGST) on August 26, 2008. Besides
GRBs, the FGST will vastly increase our knowledge
of astronomical $\gamma$-ray sources.
The historical perspective presented here will soon be
superseded by the FGST, but gives us the opportunity 
to guess what it might see.

Besides the steady diffuse glow of the Milky Way at 100 MeV 
-- GeV energies from the decay of pions formed as secondaries
in cosmic-ray collisions with dust and gas, the $\gamma$-ray sky
is pulsings from pulsars, flarings from the Sun and blazars, burstings
from GRBs, and revealing $\g$-ray enhancements from sources in the Solar system,
Galaxy, and  beyond the Galaxy at sites of cosmic-ray production 
or dark-matter annihilation.

The FGST will join an ongoing revolution in high-energy astronomy, 
including ground-based air and water Cherenkov telescopes VERITAS, 
HESS, Cangaroo III,
MAGIC-2, and Milagro and its successor HAWC.
At GeV energies, AGILE (with its super-AGILE X-ray detector), will
be joined by the FGST with its GBM and LAT. This is in addition
to multimessenger observatories 
(including Auger, KM3NeT, HiRes, LIGO, LISA, ANITA) and X-ray 
telescopes (e.g., Chandra, 
XMM, RXTE, Suzaku\dots.) ({\it Exercise:}
Report on various observatories.)

\subsection{EGRET and Fermi/GLAST}
The Energetic Gamma Ray Experiment Telescope (EGRET) 
on CGRO was most sensitive between $\approx 100$ MeV -- 5 GeV 
with a field-of-view equal to about 1/24$^{th}$ of the full sky.
Its maximum effective area between 100 MeV and 1 GeV was about
1200 cm$^2$, and its
point-spread function was $\approx 5.7^\circ$ at 100 MeV, improving
$\propto E^{-1/2}$ at higher energies. For a nominal two-week observing
period, EGRET reached integral fluxes $\gtrsim 15\times 10^{-8}$
ph($>100$ MeV) cm$^{-2}$ s$^{-1}$ for high-latitude point sources. 
The FOV is defined as the subtended solid angle swept 
from the zenith to the angle 
where the effective area is one-half of on-axis effective area.

EGRET provided tantalizing clues to $\gamma$-ray emission 
from GRBs, including GRB 940217 with a tail of 100 MeV -- GeV
emission to $\sim 93$ minutes after the GRB trigger 
(Hurley et al.\ 1994).  A 20 GeV
photon appeared after the end of Earth occultation.
Forming crude spectra during prompt phase and afterglow phase
shows clear evidence for a second, high energy emission component. 
Moreover, $\sim 100$ MeV $\gamma$ rays are found during peak pulses. 

Not only GRB 940217, but the superbowl GRB 930131, which could
be a SGRB, displayed hard emission after the end of the 
prompt phase. The prototype Milagro experiment presented evidence 
for multi-TeV emission from GRB 9704171A, which can best be confirmed
if new members of this class were detected.

Most important for expectations with the FGST 
is to consider results for five GRBs from the 
EGRET Spark Chamber. The average spectrum of 
high energy GRB emission within 200 seconds 
of BATSE trigger for GRB 940301, GRB 940217, GRB 930131, GRB 910601, GRB 940503 is well fit with 
a differential power law photon spectrum
with  index $1.95^{+0.25}_{-0.4}$  (Dingus et al.\ 1998).
Seven GRBs were detected with EGRET 
either during the prompt sub-MeV burst, 
or after sub-MeV emission has decayed away.
The ratios of the $100$ MeV -- 5 GeV EGRET fluence
to $>20$ keV BATSE fluence are typically a few percent to a few
tens of percent. (Note that these GRBs were all observed early in 
the EGRET mission before the spark chamber gas was degraded. 
Silicon strip detector technology in the FGST avoids the use of consumables.)

The Total Absorption Calorimeter on EGRET scintillated in 
response to very bright GRB events, even for off-axis GRBs. 
Joint BATSE/EGRET TASC analysis 
discovered  anomalous $\gamma$-ray emission components in 
GRB 941017 not easily explained with the leptonic blast-wave 
model (Gonz{\'a}lez et al.\ 2003). 

The FGST, as already noted, consists of two telescopes,
the Large Area Telescope, the LAT, and the Fermi GLAST Burst Monitor,
GBM. 
The sensitive energy range of the LAT is between 50 MeV
and 100 GeV (self-vetoing backsplash limited the highest energies
of EGRET to 5 GeV), with a FOV of 1/5$^{th}$ of the full sky,
and a PSF of $\approx 3.5^\circ$ ($0.55^\circ$) 
at 100 MeV (1 GeV). Its on-axis
effective area for GeV photons is $\approx 9000$ cm$^{-2}$, and 
the nominal observing strategy is to scan the sky every three hours.
The combined effective area and smaller FOV means that the FGST can 
reproduce EGRET's one-year capabilities within about 4 days, and will reach one-year
detection thresholds of $\approx 0.4\times 10^{-8}$ ph($> 100$
MeV)/cm$^{2}$-s. The GBM can, to first order, be considered 
comparable to a slightly smaller BATSE, detecting $\approx 200$ GRBs per year, 
but with better sensitivity between $\approx 1$ -- 30 MeV due to 
the BGO (Bismuth Germanate) scintillator, as well as sensitive down to $\approx 8$ keV.

The effective area for the the FGST telescope depends on the observing
mode, but naive comparison of effective areas and point spread
functions reveals how superior the FGST will be with
respect to EGRET. 
(For the latest Fermi LAT 
instrument performance, google ``GLAST LAT Performance.")
Thin and thick sections refer to different thicknesses of conversion
layers in the LAT tracker to optimize for effective area 
(thick layers) or direction (thin layers). 
Different predictions
about FGST LAT detection of GRBs can be made by scaling from
the relative ratios of fluences in BATSE to EGRET for the 5 spark 
chamber GRBs.  About 20 GRBs full sky
per year with more than 10 ($> 100$ MeV) $\gamma$ rays, or about
four per year in the LAT FOV with more than 5 ($> 100$ MeV) $\gamma$ rays 
(excluding autonomous slewing maneuvers), are predicted (Le \& Dermer 2008).  
Only about 1 GRB with more 
than 100 ($> 100$ MeV) $\gamma$ rays is predicted through the five-year 
nominal Fermi/GLAST lifetime, but presence of second components or new classes 
of GRBs should improve detection rate.
 
FGST data is proprietary until one year after the start of 
Phase I, the first year of science operations. 
Even during
the first year, the GLAST team is obligated to release light curves
and spectral data as soon as practical for $\gamma$-ray transients
that exceed a flux of $\approx 200\times 10^{-8}$ ph($> 100$
MeV)/cm$^{2}$-s, which should occur at the rate of once every week or
so. 
On top of that, the light curves and spectral behavior
of some 24 sources of interest will be released to the community.

Analysis of EGRET data to search for point and extended points
of radiation requires, especially at low ($|b|<10^\circ$) 
Galactic latitudes, a diffuse model for cosmic-ray interactions in 
the Galaxy. 
The third EGRET catalog (Hartman et al.\ 1999) lists
271 $\gamma$-ray sources, including
 the single 1991 solar flare bright 
enough to be detected as a $\gamma$-ray source, 
the Large Magellanic Cloud, five pulsars, 
one probable radio galaxy detection (Cen A), and 66 high-confidence 
identifications of blazars (BL Lac objects, flat-spectrum radio quasars, 
or unidentified flat-spectrum radio sources). In addition, 27 lower 
confidence potential blazar identifications are noted. Finally, the 
catalog contains 170 sources not yet identified firmly with known objects.

Surprises and important results are expected
from FGST regarding
\begin{enumerate}
\item $\gamma$ rays from associations of high mass stars with strong stellar winds; 
\item Spectra of $\gamma$-ray emission from supernova remnants, allowing $\gamma$-ray astronomy
a final chance to answer the question of Galactic cosmic-ray origin; 
\item Origin of microquasar $\gamma$-ray emissions, whether due to 
a scaled-down microquasar jet or to particles accelerated the shocks
formed at the interface of a binary system consisting of a young pulsar
and a high-mass star;
\item Galactic astronomy, using other data bases (e.g., WMAP) to 
search for gaseous components in the Galaxy that are illuminated by 
cosmic-ray interactions; 
\item Search for $\gamma$ rays from normal, starburst, and IR luminous 
galaxies;
\item $\gamma$ rays from cosmic rays energized by structure formation 
shocks in clusters of galaxies; 
\item $\gamma$ rays from dark matter annihilation; 
\item Blazars and their misaligned radio galaxy counterparts.
\end{enumerate}

\subsection{AGN Studies with the Fermi Gamma-ray Space Telescope}

The standard blazar model features collimated ejection of relativistic 
plasma from supermassive black holes. The relativistic motion accounts 
for lack of $\g\g$ attenuation, as in the case of GRBs. Other evidence for relativistic 
outflows include superluminal motion, super-Eddington luminosities,
high-energy beamed $\gamma$ rays made in Compton or photo-hadronic processes.
There is a considerable difference in the environments of GRBs and blazars, 
for example, the intense external radiation field from broad line-region gas
in FSRQs.

The redshift distribution of EGRET blazars leads to 
model predictions for the FGST.
Connections between different
types of supermassive black holes with jets of radio and $\gamma$-ray
emitting plasma 
can be made based on data with the FGST, 
including tests of the blazar main sequence, and studies of
black-hole engine and jet physics. 
FGST observations with correlated
multiwavelength campaigns will provide important data for modeling
studies, and for searching for anomalous blazar $\gamma$-ray emission 
components, including orphan flares and hadronic $\gamma$-ray signatures. 
Long straight radio and X-ray jets, like in the FR2 radio galaxy
Pictor A, could reveal UHECR acceleration in blazars through neutral beam
processes. 
Acceleration of UHECRs in blazars, like in GRBs, will be 
most decisively demonstrated with neutrino detection. 
The detection
of one or two $\sim $PeV neutrinos from a blazar during flaring
conditions will overturn  our thinking about how radio lobes are formed, 
because a decaying neutrons and attenuated UHE $\gamma$-rays can power
the knots, hot spots, and lobes of a radio galaxy. Detection of pair
halos from nearby radio galaxies would give evidence in favor of the 
UHECRs made in GRBs.

The superposition of $\gamma$-rays formed by various sources throughout
cosmic time produce an unresolved $\gamma$-ray background intensity. 
Only about $\sim 10$ -- 20\% of the diffuse background can be made
by FSRQ and BL Lac blazars, based on analysis of the EGRET results 
(Dermer 2007). Other source classes that can make up the diffuse
extragalactic $\gamma$-ray background include 
\begin{enumerate}
\item Star-forming galaxies;
\item Starburst galaxies;
 \item  Pulsars;
\item Galaxy cluster shocks; and
\item Dark matter annihilation.
\end{enumerate}
Although GRBs are momentarily very bright, they are so rare that their
contribution to the diffuse background is small. 
({\it Problem:} Construct fully analytic or semi-analytic
[solution in quadrature, i.e., single integral] models of 
various source classes.)


\subsection{Concluding Remarks}

Some important problems in GeV $\gamma$-ray astronomy
that will soon be opened for study in view of anticipated FGST discoveries
include 

\begin{enumerate}
\item Particle acceleration theory;
\item Origin of the galactic cosmic rays;
\item Jet physics, and the differences between radio/$\g$-ray black-hole sources;
\item Blazar demographics;
\item Search for hadronic $\gamma$-ray emission components and the sources of 
 UHECRs; and the 
\item Origin of the diffuse/unresolved $\gamma$-ray background.
\end{enumerate}

In view of the Fermi Gamma ray Space Telescope for GRB studies, we can expect a new golden era.

\acknowledgments The first author would like to thank Professors 
Gustavo Romero, Paula Benaglia, and Ileana Andruchow for 
the kind invitation and the opportunity to lecture
at the First La Plata School in Astrophysics and Geophysics, 
La Plata, Argentina, March 10 -- 14, 2008. 
Special thanks are owed to Professor Pablo M. Cincotta for hospitality, 
to Dr.\ Kevin Hurley for extensive use of material from his 
lectures given at the International School for Space Science, 
Sept. 12-16, 2005, L'Aquila, Italy and, especially, to the
students at the First La Plata International 
School on Astronomy and Geophysics for a 
truly enjoyable experience in  Argentina. I would especially like to thank
Ignacio Francisco Ranea Sandoval and Andrea Cesarini for their particular
interest. The research of CDD is supported by the NASA GLAST 
Interdisciplinary Scientist Program and the Office of Naval Research.

\end{document}